\def\nk{n_{\rm b}}

\def\Pb{P_{\rm b}}

\def\rfr#1{Equation\,(\ref{#1})}
\def\rfrs#1#2{Equations\,(\ref{#1})-(\ref{#2})}
\def\Rfr#1{Equation\,(\ref{#1})}

\def\derp#1#2{\rp{\partial{#1}}{\partial{#2}}}
\def\dert#1#2{\frac{{{\textrm{d}}}{#1}}{{{\textrm{d}}}{#2}}}

\def\virg#1{``#1"}

\def\eqi{\begin{equation}}
\def\eqf{\end{equation}}
\def\eqia{\begin{eqnarray}}
\def\eqfa{\end{eqnarray}}

\def\rp#1#2{{#1\over#2}}
\def\lb#1{\label{#1}}

\def\bds#1{\boldsymbol{#1}}

\def\ton#1{\left(#1\right)}
\def\qua#1{\left[#1\right]}
\def\grf#1{\left\{#1\right\}}

\documentclass[onecolumn]{aastex}
\usepackage{morefloats}
\usepackage[title]{appendix}
\usepackage{hyperref}
\usepackage{booktabs}
\usepackage[table,xcdraw]{xcolor}
\usepackage{multirow}
\usepackage{rotating,tabularx}
\usepackage{float}
\usepackage{enumerate}
\usepackage{rotating}
\usepackage[polutonikogreek,english]{babel}
\usepackage{amsmath,starfont,textgreek,w-greek,wasysym}
\usepackage{amsthm}
\usepackage{amscd,lineno}
\usepackage{amssymb,dsfont}
\usepackage{graphicx,epsfig}
\usepackage{txfonts}
\usepackage{xr-hyper}

\RequirePackage{color}

\newcommand{\emaila}{lorenzo.iorio@libero.it}

\allowdisplaybreaks[1]

\begin{document}

\title{On the 2PN pericentre precession in the general theory of relativity and the recently discovered fast orbiting S-stars in Sgr A$^\ast$
}

\shortauthors{L. Iorio}

\author{Lorenzo Iorio\altaffilmark{1} }
\affil{Ministero dell'Istruzione, dell'Universit\`{a} e della Ricerca
(M.I.U.R.)-Istruzione
\\ Permanent address for correspondence: Viale Unit\`{a} di Italia 68, 70125, Bari (BA),
Italy}

\email{\emaila}

\begin{abstract}
Recently, the secular pericentre precession was analytically computed  to the second post-Newtonian (2PN) order by the present author with the Gauss equations in terms of the osculating Keplerian orbital elements in order to obtain closer contact with the observations in astronomical and astrophysical scenarios of potential interest. A discrepancy with previous results by other authors was found. Moreover, some of such findings by the same authors were deemed as mutually inconsistent. In this paper, it is demonstrated that, in fact, \textcolor{black}{some} calculational errors plagued the most recent calculation by the present author. They are explicitly disclosed and corrected. As a result, all the examined approaches  mutually agree yielding the same analytical expression for the total 2PN pericentre precession once the appropriate conversions from the adopted parameterizations are made. It is also shown that, in future, it may become measurable, at least in principle, for some of the recently discovered short-period S-stars in Sgr A$^\ast$ like S62 and S4714.
\end{abstract}

keywords{
general relativity and gravitation;  celestial mechanics
}
\section{Introduction}
The analytical calculation of the secular 2PN\footnote{For an overview of Post-Newtonian (PN) theory and its applications, see, e.g., \citet{2003Blanchet,2014LRR....17....4W,2018tegp.book.....W} and references therein.} pericentre\footnote{The precession of the pericentre is one of the post-Keplerian (pK) parameters which are used in testing gravitational theories in astrophysical binary systems containing at least one compact object; see, e.g., \citet{2019ApJ...874..121Z}.} precession $\dot\omega^\mathrm{2PN}$ of a gravitationally bound two-body system made of two mass monopoles $M_\mathrm{A},\,M_\mathrm{B}$  with the perturbative Gauss equations for the variation of the osculating Keplerian orbital elements \citep[e.g.][]{2011rcms.book.....K,SoffelHan19} was the subject of \citet{2020Univ....6...53I}.  For the sake of simplicity, the test particle limit will be considered in most of the paper.
In the following, $c$ is the speed of light in vacuum, $\mu\doteq GM$ is the gravitational parameter of the primary whose mass is $M$, $G$ is the Newtonian constant of gravitation, $\mathbf{v},\,\mathrm{v}_r,\,\mathrm{v}$ are the test particle's velocity, radial velocity and speed, respectively, $r$ is the test particle's distance from the primary, $\bds{\hat{r}}$ is the position unit vector of the test particle with respect to the primary, $f_0,\,a,\,e$ are \citep{1994ApJ...427..951K}
the osculating numerical values of the true anomaly, semimajor axis and eccentricity, respectively, at the same arbitrary moment of time $t_0$, and $\nk=\sqrt{\mu/a^3}$ is the osculating mean motion.

The expression for the total 2PN pericentre precession derived by \citet{2020Univ....6...53I} consists of the sum of three contributions. The first one, dubbed as \virg{direct}, is  \citep[Equation\,(8)]{2020Univ....6...53I}
\eqi
\dot\omega_\mathrm{dir}^\mathrm{2PN} = \rp{\nk\,\mu^2\,\ton{28 - e^2}}{4\,c^4\,a^2\,\ton{1-e^2}^2},\lb{dire2PN}
\eqf
arising straightforwardly from the 2PN acceleration
\eqi
\bds{A}^\mathrm{2PN} = \rp{\mu^2}{c^4\,r^3}\qua{\ton{2\,{\mathrm{v}}_r^2-\rp{9\,\mu}{r}}\bds{\hat{r}} - 2\,{\mathrm{v}}_r\,{\mathbf{v}} }.\lb{acc2PN}
\eqf
There are also two further contributions, labeled as \virg{mixed} or \virg{indirect}. They account for the fact that, when the Gauss equation for the rate of change of the pericentre induced by the 1PN acceleration
\eqi
{\bds{A}}^\mathrm{1PN} = \rp{\mu}{c^2\,r^2}\,\qua{\ton{\rp{4\,\mu}{r} -\mathrm{v}^2 }\,\bds{\hat{r}} +4\,{\mathrm{v}}_r\,{\mathbf{v}} }\lb{acc1PN}
\eqf
is averaged over one orbital period $\Pb$, the latter one has to be considered as the time interval between two consecutive crossings of the moving pericentre. Moreover, also the instantaneous shifts of the other orbital elements due to \rfr{acc1PN} itself are to be taken into account when the orbital average is performed. Both such effects contribute the total pericentre precession to the 2PN level.
The first indirect effect yields \citep[Equation\,(14)]{2020Univ....6...53I}
\eqi
\dot\omega_\mathrm{indir}^\mathrm{2PN\,\ton{I}} = \rp{\nk\,\mu^2\,\ton{9 + 37\,e^2 + e^4}}{2\,c^4\,a^2\,e^2\,\ton{1-e^2}^2},\lb{indir2PNI}
\eqf
while the second indirect contribution reads \citep[Equation\,(22)]{2020Univ....6...53I}
\eqi
\dot\omega_\mathrm{indir}^{\mathrm{2PN\,\ton{II}}} = -\rp{\nk\,\mu^2\,\grf{
9 - 87\,e^2 - 136\,e^4 + 19\,e^6 - 6\,e^3\,\qua{\ton{34 + 26\,e^2}\,\cos f_0 + 15\,e\,\cos 2 f_0}}}{2\,c^4\,e^2\,a^2\,\ton{1-e^2}^3}\lb{indir2PNII}.
\eqf
Thus, the sum of \rfr{dire2PN}, \rfr{indir2PNI}, and \rfr{indir2PNII} gives the total 2PN pericentre precession
\eqi
\dot\omega_\mathrm{tot}^\mathrm{2PN} = \rp{3\,\nk\,\mu^2\,\qua{86 + 57\,e^2 - 13\,e^4 + 8\,e\,\ton{17 + 13\,e^2}\,\cos f_0 +
 60\,e^2\,\cos 2f_0}}{4\,c^4\,a^2\,\ton{1-e^2}^3}.\lb{tot2PN}
\eqf

\citet{2020Univ....6...53I} compared his results with those by \citet{1994ARep...38..104K}, who used the perturbative approach relying upon the Gauss equations as well, and those by \citet{1988NCimB.101..127D}, obtained with the Hamilton-Jacobi method. A discrepancy with such authors was found since \citet{2020Univ....6...53I} \textcolor{black}{claimed} that the total 2PN pericentre precession inferred by \citet[Equation\,(5.2)]{1994ARep...38..104K} can be cast into the form  \textcolor{black}{\citep[Equation\,(53)]{2020Univ....6...53I}}
\eqi
\dot\omega_\mathrm{tot}^\mathrm{2PN}=\rp{3\,\nk\,\mu^2\,\ton{2 + e^2 -32\,e^2\,\cos f_0}}{4\,c^4\,a^2\,\ton{1-e^2}^2}.\lb{kazzoo}
\eqf
\textcolor{black}{Actually, a typo occurred in  \citet[Equation\,(53)]{2020Univ....6...53I} since the correct expression reads
\eqi
\dot\omega_\mathrm{tot}^\mathrm{2PN}=\rp{3\,\nk\,\mu^2\,\ton{2 + e^2 -32\,e\,\cos f_0}}{4\,c^4\,a^2\,\ton{1-e^2}^2};\lb{totKoPo}
\eqf
see Section\,\ref{sec3}.
\Rfr{totKoPo} was not explicitly shown by \citet{1994ARep...38..104K}; it was mistakenly reproduced in \citet[Equation\,(44)]{Kop020}
by writing $-32\,\cos f_0$ in the numerator.
}

Furthermore, \citet{2020Univ....6...53I} claimed that \citet[Equation\,(3.12)]{1988NCimB.101..127D}, which was demonstrated to be coincident with \rfr{totKoPo} \textcolor{black}{once the aforementioned typo is taken into account}, and \citet[Equation\,(5.18)]{1988NCimB.101..127D} would be mutually inconsistent.

Here, it will be proven that, actually, a mere calculational error occurred in the derivation of $\dot\omega_\mathrm{indir}^\mathrm{2PN\,\ton{II}}$ by \citet{2020Univ....6...53I} which prevented to obtain \rfr{totKoPo} instead of the incorrect \rfr{tot2PN}. Once such an error is corrected, \textcolor{black}{ and the typo in \rfr{kazzoo} removed,} both the approaches by \citet{2020Univ....6...53I} and \citet{1994ARep...38..104K}, which differ in how obtaining just $\dot\omega_\mathrm{indir}^\mathrm{2PN\,\ton{II}}$, agree yielding the same total 2PN pericentre precession of \rfr{totKoPo}.
Moreover, it will be shown that also the alleged inconsistency of Equation\,(3.12) and Equation\,(5.18) by \citet{1988NCimB.101..127D} is, in fact, due to another error by \citet{2020Univ....6...53I}, as correctly pointed out by \citet{Kop020}. \textcolor{black}{For other examples of different parameterizations used in calculating the 2PN precession, see, e.g., \citet{Pau21}.}

To the benefit of the reader, it is noted that \citet{1988NCimB.101..127D,1994ARep...38..104K} usually dealt with the fractional pericentre advance per orbit, i.e., $\Delta\omega/2\uppi$; in order to obtain the corresponding precession, it is sufficient to multiply it by $\nk$.

The paper is organized as follows.

In Section\,\ref{sec3}, the calculational error in working out $\dot\omega_\mathrm{indir}^\mathrm{2PN\,\ton{II}}$ is explicitly disclosed and corrected\textcolor{black}{, and \rfr{totKoPo} is obtained}. Section\,\ref{sec4} is devoted to showing, independently of \citet{Kop020}, that  Equation\,(3.12) and Equation\,(5.18) of \citet{1988NCimB.101..127D} are, actually, mutually consistent yielding both the same total 2PN pericentre precession as \rfr{totKoPo}.
Some aspects of the presence of $f_0$ in \rfr{totKoPo} are discussed in Section\,\ref{sec5}.
Section\,\ref{sec6} is devoted to calculating the 2PN pericentre precession for some of the recently discovered fast orbiting S-stars \citep{2020ApJ...899...50P} in the Galactic Center (GC) at Sgr A$^\ast$.
Section\,\ref{sec7} summarizes the present findings and offers concluding remarks.

\section{Disclosing and correcting the error for $\dot\omega_\mathrm{indir}^\mathrm{2PN\,\ton{II}}$}\lb{sec3}
In  \citet{2020Univ....6...53I}, it turned out that \rfr{dire2PN} and \rfr{indir2PNI} agree with the corresponding calculation by \citet{1994ARep...38..104K}, despite such authors did neither recur to the schematization by \citet{2020Univ....6...53I} nor explicitly display their intermediate results.

Instead,  \citet{2020Univ....6...53I} realized that the discrepancy among his results and those by \citet{1994ARep...38..104K} resides in $\dot\omega_\mathrm{indir}^\mathrm{2PN\,(II)}$, i.e., in that part of the indirect precession arising from the fact that the semimajor axis and the eccentricity do change instantaneously during an orbital revolution due to \rfr{acc1PN} according to
\begin{align}
\Delta a\ton{f_0,\,f}^\mathrm{1PN} \lb{Da1PN} &= -\rp{2\,e\,\mu\,\ton{\cos f-\cos f_0}\,\qua{7 + 3\,e^2 + 5\,e\,\ton{\cos f+\cos f_0}}}{c^2\,\ton{1-e^2}^2}, \\ \nonumber \\
\Delta e\ton{f_0,\,f}^\mathrm{1PN} \lb{De1PN} &= \rp{\mu\,\ton{\cos f_0 - \cos f}\,\qua{3 + 7\,e^2 + 5\,e\,\ton{\cos f+\cos f_0}}}{c^2\,a\,\ton{1-e^2}}.
\end{align}
The calculation of $\dot\omega_\mathrm{indir}^\mathrm{2PN\,(II)}$ by \citet{1994ARep...38..104K} can be reproduced as follows \citep[pp.\,13]{2020Univ....6...53I}.
Evaluate the Gauss equation of the pericentre for a perturbing in-plane acceleration
\eqi
\dert{\omega}{f} = \rp{r^2}{\mu e}\,\qua{-A_\rho\,\cos f+\ton{1+\rp{r}{p}}\sin f\,A_\tau},\lb{gaus}
\eqf
where $p\doteq a\ton{1-e^2}$, with the radial and transverse components of the 1PN acceleration of \rfr{acc1PN}
\begin{align}
A^\mathrm{1PN}_\rho \lb{acc1PNr} & = \rp{\mu^2\,\ton{1 + e\,\cos f}^2\,\ton{3 + e^2 + 2\,e\,\cos f - 2\,e^2\,\cos 2 f}}{c^2\,a^3\,\ton{1-e^2}^3},\\ \nonumber \\
A^\mathrm{1PN}_\tau \lb{acc1PNt} & = \rp{4\,e\,\mu^2\,\ton{1 + e\,\cos f}^3\,\sin f}{c^2\,a^3\,\ton{1-e^2}^3}.
\end{align}
Then, make the replacement
\begin{align}
a& \rightarrow a + \Delta a\ton{f_0,\,f}^\mathrm{1PN}, \\ \nonumber \\
e& \rightarrow e + \Delta e\ton{f_0,\,f}^\mathrm{1PN},
\end{align}
by means of \rfrs{Da1PN}{De1PN},
expand $\mathrm{d}\omega/df$ to the order of $\mathcal{O}\ton{c^{-4}}$, and integrate the resulting expression
\begin{align}
\left.\dert{\omega}{f}\right|^\mathrm{2PN\,\ton{II}}_\mathrm{indir} \nonumber \lb{matoo} &= -\rp{\mu^2\,\ton{\cos f -\cos f_0}}{2\,c^4\,a^2\,e^2\,\ton{1-e^2}^2}\,\grf{e\,\qua{15 - 43\,e^2 + 5\,\ton{3 + 17\,e^2}\,\cos 2f} + \right. \\ \nonumber \\
& \left. + 2\,\cos f\,\qua{9 + 48\,e^2 - e^4 + 5\,e\,\ton{3 + e^2}\,\cos f_0}}
\end{align}
from $f_0$ to $f_0+2\uppi$. By dividing the result by $1/\Pb=\nk/2\uppi$, the following formula is obtained
\eqi
\dot\omega^\mathrm{2PN\,\ton{II}}_\mathrm{indir} = \rp{\nk\,\mu^2\,\ton{-9 - 48\,e^2 + e^4 - 48\,e^3\,\cos f_0}}{2\,c^4\,a^2\,e^2\,\ton{1-e^2}^2},\lb{koppo}
\eqf
so that the sum of \rfr{indir2PNI} and \rfr{koppo} gives the total 2PN indirect precession \citep[Equation\,(55)]{2020Univ....6...53I}
\eqi
\dot\omega^\mathrm{2PN}_\mathrm{indir} = \rp{\nk\,\mu^2\,\ton{-11 + 2\,e^2 - 48\,e\,\cos f_0}}{2\,c^4\,a^2\,\ton{1-e^2}^2},\lb{strunz}
\eqf
which is, actually, correct.
It turns out that \textcolor{black}{summing} \rfr{strunz} to \rfr{dire2PN}  yields just the total 2PN precession of \rfr{totKoPo}.

\citet[pag.\,5]{2020Univ....6...53I} followed another approach in calculating $\dot\omega_\mathrm{indir}^\mathrm{2PN\,\ton{II}}$. In the specific case of the pericentre and of \rfr{acc1PN}, starting from \rfr{gaus}, calculated with \rfrs{acc1PNr}{acc1PNt}, the net 2PN pericentre shift per orbit due to the 1PN instantaneous variations  of \rfrs{Da1PN}{De1PN} is worked out as
\eqi
\Delta\omega_\mathrm{indir}^\mathrm{2PN\,\ton{II}} = \int_{f_0}^{f_0+2\uppi}\qua{\derp{\ton{\mathrm{d}\omega/\mathrm{d}f}}{a}\,
\Delta a\ton{f_0,\,f}^\mathrm{1PN} + \derp{\ton{\mathrm{d}\omega/\mathrm{d}f}}{e}\,
\Delta e\ton{f_0,\,f}^\mathrm{1PN}}\,\mathrm{d}f.\lb{gro}
\eqf
A calculational error\footnote{To be more specific, $\mu^2$ entering \rfrs{acc1PNr}{acc1PNt} was expressed in $\mathrm{d}\omega/\mathrm{d}f$ as $\nk^4\,a^6$, thus altering the partial derivative of $\mathrm{d}\omega/\mathrm{d}f$ with respect to $a$.} in the first addend of \rfr{gro} yielded the wrong result of \rfr{indir2PNII}. After correcting it, it is possible to show that the function to be integrated in \rfr{gro} agrees with \rfr{matoo}. Thus, \rfr{koppo} can be correctly obtained also with the method for calculating $\dot\omega_\mathrm{indir}^\mathrm{2PN\,\ton{II}}$ used by \citet{2020Univ....6...53I}.

By repeating the calculation by \citet{2020Univ....6...53I}, corrected for the aforementioned error, one obtains, for the full two-body system,
\eqi
\dot\omega_\mathrm{indir}^\mathrm{2PN} =  \rp{\nk\,\mu^2\,\qua{-44 + 8\,\nu\,\ton{-8 + 7\,\nu} + e^2\,\ton{8 + 39\,\nu + 48\,\nu^2} +
 96\,e\,\ton{-2 +\,\nu}\,\cos f_0}}{8\,c^4\,a^2\ton{1-e^2}^2}\lb{mixtot}
\eqf
which, \textcolor{black}{summed} to \citet[Equation\,(32)]{2020Univ....6...53I}, returns
\eqi
\dot\omega_\mathrm{tot}^\mathrm{2PN} = \rp{3\,\nk\,\mu^2\,\qua{2 - 4\,\nu + e^2\,\ton{1 + 10\,\nu} + 16\,e\,\ton{-2 + \nu}\,\cos f_0}}{4\,c^4\,a^2\ton{1-e^2}^2}.\lb{finis}
\eqf
In \rfrs{mixtot}{finis}, it is
\begin{align}
\nu &\doteq \rp{M_\mathrm{A}\,M_\mathrm{B}}{\ton{M_\mathrm{A}+M_\mathrm{B}}^2}, \\ \nonumber \\
\mu &\doteq G\ton{M_\mathrm{A} + M_\mathrm{B}}.
\end{align}
\textcolor{black}{ In the limit $\nu\rightarrow 0$, \rfr{finis} agrees with \rfr{totKoPo}.}

\textcolor{black}{
It is, now, explicitly shown that \citet[Equation\,(5.2)]{1994ARep...38..104K} coincides with our \rfr{totKoPo} after the  traduction from one parameterization to another is properly carried on.
The starting point is, in the test particle limit, \citep[Equation\,(5.2)]{1994ARep...38..104K}
\eqi
\rp{\Delta\omega_\mathrm{tot}}{2\uppi}=
\rp{3\,\mu}{c^2\,k_1\,\ton{1-k_2^2}}\qua{1+\rp{3\,\mu}{4\,c^2\,k_1\,\ton{1-k_2^2}}-\rp{\mu}{4\,c^2\,k_1}}.\lb{KPK}
\eqf
In it, $k_1,\,k_2$ are the constants of integration\footnote{\citet{1994ARep...38..104K}, used the notation $a_0,\,e_0$ instead of $k_1,\,k_2$.}  of the solutions of the Gauss equations for the semimajor axis and the eccentricity to the 1PN level, to be determined with the initial conditions at $t = t_0$.
They can be obtained, e.g., by evaluating \citet[Equations\,(4.5)-(4.6)]{1994ARep...38..104K} at $t=t_0$ by replacing $f$, i.e. $V$ in the notation by \citet{1994ARep...38..104K}, with $f_0$ and by recalling that, in our notation, $a_0\rightarrow k_1,\,e_0\rightarrow k_2$. Moreover, $a,\,e$ are, for us, the osculating numerical values of the semimajor axis and eccentricity, respectively, at the same arbitrary instant $t_0$; thus, $a(t_0),\,e(t_0)$ in the left-hand side of \citet[Equations\,(4.5)-(4.6)]{1994ARep...38..104K} are just our $a,\,e$ here. Thus, in the limit $\nu\rightarrow 0$, one gets\textcolor{black}{\footnote{\textcolor{black}{A calculational error occurred in \citet[Equations\,(51)-(52)]{2020Univ....6...53I} for $k_1,\,k_2$.}}}
\begin{align}
k_1 \nonumber \lb{KA1} & = \rp{a}{1+\rp{e\,\mu}{c^2\,a\,\ton{1-e^2}^2}\,\qua{\ton{- 14 - 6\,e^2}\,\cos f_0 -5\,e\,\cos 2f_0  }}\simeq \\ \nonumber \\
&\simeq a + \rp{e\,\mu}{c^2\,\ton{1-e^2}^2}\,\qua{\ton{14 + 6\,e^2}\,\cos f_0 + 5\,e\,\cos 2f_0}, \\ \nonumber \\
k_2 \nonumber \lb{KA2} & = \rp{e}{1 + \rp{\mu}{c^2\,e^2\,a\,\ton{1-e^2}}\,\qua{\ton{- 3 - 7\,e^2}\,e\,\cos f_0 -\rp{5}{2}\,e^2\,\cos 2f_0}}\simeq \\ \nonumber \\
&\simeq e + \rp{\mu}{2\,c^2\,a\,\ton{1-e^2}}\,\qua{\ton{6 + 14\,e^2}\,\cos f_0 + 5\,e\,\cos 2f_0}.
\end{align}
By inserting \rfrs{KA1}{KA2} into \rfr{KPK}, multiplied by $\nk$, and expanding the resulting expression in powers of $c^{-1}$ to the order $c^{-4}$, one gets just the sum of \rfr{totKoPo} and of the 1PN precession
\eqi
\dot\omega^\mathrm{1PN} = \rp{3\,\nk\,\mu}{c^2\,a\,\ton{1-e^2}}.\lb{konzo}
\eqf
}

As far as Mercury is concerned,
for which it is \citep[Fig.\,1]{2020Univ....6...53I}
\eqi
\dot\omega^\mathrm{2PN}_\mathrm{dir} = 2.6\,\mu\mathrm{as\,cty}^{-1},
\eqf
where $\mu\mathrm{as\,cty}^{-1}$ stands for microarcseconds per century, \rfr{strunz} yields
\eqi
-4\,\mu\mathrm{as\,cty}^{-1}\leq\dot\omega_\mathrm{indir}^\mathrm{2PN}\leq -0.2\,\mu\mathrm{as\,cty}^{-1},\lb{mercu}
\eqf
for $0^\circ\leq f_0\leq 360^\circ$; \rfr{mercu} corrects \citet[Equation\,(24)]{2020Univ....6...53I}.

For the double pulsar PSR J0737-3039A/B, for which it is \citep[Equation\,(33)]{2020Univ....6...53I}
\eqi
\dot\omega_\mathrm{dir}^\mathrm{2PN} = 0.00019^\circ\,\mathrm{yr}^{-1},
\eqf
from \rfr{mixtot}, it turns out
\eqi
-0.00022^\circ\,\mathrm{yr}^{-1}\leq\dot\omega_\mathrm{indir}^\mathrm{2PN}\leq -0.00013^\circ\,\mathrm{yr}^{-1}\lb{doppia}
\eqf
for $0^\circ\leq f_0\leq 360^\circ$.
For the Hulse-Taylor binary pulsar PSR B1913+16, for which it is \citep[Equation\,(35)]{2020Univ....6...53I}
\eqi
\dot\omega_\mathrm{dir}^\mathrm{2PN} = 0.000038^\circ\,\mathrm{yr}^{-1},
\eqf
\rfr{mixtot} yields
\eqi
-0.00009^\circ\,\mathrm{yr}^{-1}\leq\dot\omega_\mathrm{indir}^\mathrm{2PN}\leq 0.000034^\circ\,\mathrm{yr}^{-1}\lb{HuTa}
\eqf
for $0^\circ\leq f_0\leq 360^\circ$.
\Rfr{doppia} and \rfr{HuTa} correct \citet[Equations\,(48)-(49)]{2020Univ....6...53I}.

For the supermassive binary black hole in OJ 287,
for which it is \citep[pag.\,10]{2020Univ....6...53I}
\eqi
\dot\omega_\mathrm{dir}^\mathrm{2PN}=11.0^\circ\,\mathrm{cty}^{-1},
\eqf
\rfr{mixtot} returns an indirect 2PN perinigricon precession ranging within
\eqi
-33.4^\circ\,\mathrm{cty}^{-1}\leq\dot\omega_\mathrm{indir}^\mathrm{2PN}\leq 17^\circ\,\mathrm{cty}^{-1}\lb{OJ}
\eqf
for $0^\circ\leq f_0\leq 360^\circ$. \Rfr{OJ} corrects the figures yielded in \citet[pag.\,12]{2020Univ....6...53I}. In retrospect, they should  have been a wake-up call concerning the validity of \citet[Equation\,(47)]{2020Univ....6...53I} since the reported maximum value of $516^\circ\,\mathrm{cty}^{-1}$ is even larger than the 1PN precession itself amounting to \citep[Equation\,(37)]{2020Univ....6...53I} $\dot\omega^\mathrm{1PN} = 206.8^\circ\,\mathrm{cty}^{-1}$.

The discussion in \citet{2020Univ....6...53I} concerning the measurability of the 2PN pericentre precessions of Mercury and of the binary pulsars will not be repeated here.
%
%
%
%
%
\section{Correcting the error for $e_\mathrm{T}$}\lb{sec4}
As correctly pointed out by \citet{Kop020}, \citet{2020Univ....6...53I} erroneously claimed that Equation\,(3.12) of \citet{1988NCimB.101..127D}
\eqi
\rp{\Delta\omega_\mathrm{tot}^\mathrm{2PN}}{2\uppi} = \rp{3}{c^2\,h^2}\,\qua{1+\ton{\rp{5}{2}-\nu}\rp{E}{c^2} +\ton{\rp{35}{4} -\rp{5}{2}\nu }\rp{1}{c^2\,h^2} },\lb{DS1}
\eqf
and Equation\,(5.18) of \citet{1988NCimB.101..127D}
\begin{align}
\rp{\Delta\omega_\mathrm{tot}^\mathrm{2PN}}{2\uppi} \nonumber  &= \rp{3\ton{\mu\,n}^{2/3}}{c^2\,\ton{1-e^2_\mathrm{T}}}\,
\qua{1 + \rp{\ton{\mu\,n}^{2/3}}{c^2\,\ton{1-e^2_\mathrm{T}}}\,\ton{\rp{39}{4}\,x_\mathrm{A}^2 +\rp{27}{4}\,x_\mathrm{B}^2 + 15\,x_\mathrm{A}\,x_\mathrm{B}} -\right.\\ \nonumber \\
&\left.- \rp{\ton{\mu\,n}^{2/3}}{c^2}\,\ton{\rp{13}{4}\,x_\mathrm{A}^2 +\rp{1}{4}\,x_\mathrm{B}^2 + \rp{13}{3}\,x_\mathrm{A}\,x_\mathrm{B}} }\lb{DS2}
\end{align}
would be mutually inconsistent after being expressed in terms of $a,\,e,\,f_0$.
In \rfrs{DS1}{DS2}, $h$ and $E$ are the coordinate-invariant, reduced orbital angular momentum and energy, respectively,
\begin{align}
x_\mathrm{A} &\doteq \rp{M_\mathrm{A}}{M_\mathrm{A} + M_\mathrm{B}},\\ \nonumber \\
x_\mathrm{B} & \doteq \rp{M_\mathrm{B}}{M_\mathrm{A} + M_\mathrm{B}}=1-x_\mathrm{A},
\end{align}
$n$ is the PN mean motion \citep{1985AIHS...43..107D},
and $e_\mathrm{T}$ is one of the several Damour-Deruelle (DD) parameters \citep{1986AIHS...44..263D}.
More precisely, in \citet[pp.\,14-15]{2020Univ....6...53I}, it was correctly demonstrated that \rfr{DS1} yields \rfr{totKoPo}, \textcolor{black}{once the typo of \rfr{kazzoo} is corrected}. On the other hand, in \citet[pag.\,15]{2020Univ....6...53I}, it was erroneously claimed that \rfr{DS2} could not reduce to \rfr{totKoPo}\textcolor{black}{, up to the typo in \rfr{kazzoo}}.
The error consists of the fact that \citet{2020Univ....6...53I} confused  $e_\mathrm{T}$ \citep{1986AIHS...44..263D} entering \rfr{DS2} with $e_t$, another member of the DD parameterization \citep{1985AIHS...43..107D}.

Instead, it is \citep[pag.\,272]{1986AIHS...44..263D}
\eqi
e_\mathrm{T} = e_t\ton{1+\delta} + e_{\theta} - e_r\lb{ereT}.
\eqf
The parameters entering \rfr{ereT} are defined as
%
%
\citep[Equation\,(3.8\,b)]{1985AIHS...43..107D}
\eqi
e_t=\rp{e_R}{1+\rp{\mu}{c^2\,a_R}\ton{4-\rp{3}{2}\,\nu}},\lb{eti}
\eqf
\citep[Equation\,(20)]{1986AIHS...44..263D}
\eqi
\delta = \rp{\mu}{c^2 a_R}\ton{x_\mathrm{A}x_\mathrm{B}+ 2x_\mathrm{B}^2},
\eqf
\citep[Equation\,(4.13)]{1985AIHS...43..107D}
\eqi
e_\theta = e_R\,\ton{1+\rp{\mu\,\nu}{2\,c^2\,a_R}},
\eqf
\citep[Equation\,(6.3\,b)]{1985AIHS...43..107D}
\eqi
e_r = e_R\,\qua{1 - \rp{\mu}{2\,c^2\,a_R}\,\ton{x_\mathrm{A}^2 - \nu}}.\lb{er}
\eqf
In \rfrs{eti}{er}, the DD \virg{semimajor axis} of the relative motion\footnote{In \citet{2020Univ....6...53I}, it is designed as $a_r$: in fact, such a choice  may be confusing since, in \citet[Equation\,(6.3\,a)]{1985AIHS...43..107D}, such a quantity is meant as $a_r = M_\mathrm{B}\,a_R/\ton{M_\mathrm{A} + M_\mathrm{B}}$.}
$a_R$ can be expressed as \citep[Equation\,(66)]{2020Univ....6...53I}
\begin{align}
4\,\ton{1-e^2}^2\,a_R \nonumber \lb{aerre}& = 4\,\grf{a\,\ton{1 - e^2}^2 - \rp{\mu}{c^2}\,\qua{-3 + \nu + e^4\,\ton{1 + 2\,\nu} + e^2\,\ton{-13 + 7\,\nu}}} + \\ \nonumber \\
\nonumber & + e\,\rp{\mu}{c^2}\,\grf{\qua{56 + e^2\,\ton{24 - 31\,\nu} - 24\,\nu}\,\cos f_0 + \right. \\ \nonumber \\
&+\left. e\,\qua{4\,\ton{5 - 4\,\nu}\,\cos 2 f_0 - e\,\nu\,\cos 3 f_0}},
\end{align}
and the DD \virg{eccentricity} $e_R$ is given by\footnote{In \citet{2020Univ....6...53I}, it is denoted as $e_r$, but, in view of \rfr{er}, such a choice is misleading. } \citet[Equation\,(67)]{2020Univ....6...53I}
\begin{align}
8\,a\,\ton{-1+e^2}\,e_R \nonumber \lb{eerre}& = 4\,e\,\grf{2\,a\,\ton{-1 + e^2} + \rp{\mu}{c^2}\,\qua{-17 + 6\,\nu + e^2\,\ton{2 + 4\,\nu}}} + \\ \nonumber \\
\nonumber &+ \rp{\mu}{c^2}\,\grf{\qua{8\,\ton{-3 + \nu} + e^2\,\ton{-56 + 47\,\nu}}\,\cos f_0  + \right. \\ \nonumber \\
&+\left. e\,\qua{4\,\ton{-5 + 4\,\nu}\,\cos 2 f_0  + e\,\nu\,\cos 3 f_0}}.
\end{align}
It turns out that, using \rfrs{ereT}{eerre} and
\citep[Equation\,(3.7)]{1985AIHS...43..107D}
\eqi
n = \sqrt{\rp{\mu}{a_R^3}}\qua{1 + \rp{\mu}{2\,c^2\,a_R}\ton{-9+\nu}}
\eqf
in \rfr{DS2} and expanding all to the order of $\mathcal{O}\ton{c^{-4}}$
removes the previously mentioned alleged discrepancy. Indeed, now, the corresponding 2PN precession can be cast just into the form of \rfr{totKoPo}, inasmuch the same way as \rfr{DS1} did.
%
%
%
%
%
%
%
%
%
%
%
\section{Some considerations about the occurrence of $f_0$ in the 2PN pericentre precession}\lb{sec5}
The fact that $f_0$ enters \rfr{totKoPo} through \rfr{koppo} appears to be likely a general feature of that part of the long-term orbital precessions arising from the inclusion of the instantaneous shifts of all the Keplerian orbital elements in performing the average of the right-hand-side of the Gauss equation of any of them over an orbital period. Indeed, such a feature explicitly occurs also with the interplay of, e.g., the Newtonian acceleration due to the primary's quadrupole mass moment $J_2$ and \rfr{acc1PN} in calculating  \textcolor{black}{(}part of\textcolor{black}{)} the indirect rates of change of the Keplerian orbital elements of the order of $J_2/c^2$ by means of the Gauss equations, as done by\footnote{\textcolor{black}{A terminological confusion may arise if one does not properly note that \citet{2014PhRvD..89d4043W} used the adjective \virg{mixed} to denote the PN-quadrupole acceleration of the order of $J_2/c^2$ \citep{1988CeMec..42...81S,Sof89,1991ercm.book.....B} entering the equations of motion. Moreover, \citet{2014PhRvD..89d4043W} dubbed \virg{cross-term effects} the contributions to the overall orbital precessions labeled as \virg{indirect}, or \virg{mixed}, by \citet{2015IJMPD..2450067I} and here.}} \citet{2014PhRvD..89d4043W,2015IJMPD..2450067I}.

It should be remarked that \rfr{totKoPo} does not pretend, by no means, to contain any new physics with respect to what was obtained earlier, with different computational approaches and different parameterizations, by  \citet{1988NCimB.101..127D} and \citet{1994ARep...38..104K} in terms of
first integrals of motion like $E$ and $h$. Indeed, as it was shown here, Equation\,(5.2) by \citet{1994ARep...38..104K}, and Equation\,(3.12) and Equation\,(5.18) by \citet{1988NCimB.101..127D} agree with \rfr{totKoPo}. The appearance of $f_0$ in \rfr{totKoPo} is of a gauge nature, in that it does not influence the values of the integrals of motion, for a fixed orbit. It should be remembered that, after all, in any practical data analysis of specific astronomical and astrophysical systems of interest, an underlying choice of a given parameterization and gauge is always unavoidably made. In most cases, it
generally corresponds to that adopted in the present study\textcolor{black}{, as per IAU recommendations \citep{2003AJ....126.2687S},} which is just targeted to practical people aiming to designing and/or performing tests using, say, Mercury, exoplanets close to their parent stars, and S-stars in Sgr A$^\ast$.
\textcolor{black}{It should be also stressed that \rfr{totKoPo} and \rfr{finis} should not be thought as something to be used in actual tests by straightforwardly comparing them to some directly measured pericentre precessions since, depending on the specific astronomical scenario at hand, different approaches are practically used in reducing the observations in order to perform tests of gravitational theories. For example, in Solar System analyses, GTR is tested by modeling the propagation of electromagnetic waves, the dynamics of material bodies and the functioning of  measuring devices in a PN framework to a given order, and estimating, in a least-square sense, some dedicated parameters by fitting such models to a whole range of observations of different kinds collected over a century or so.
Instead, the data of binary pulsars are reduced with a timing formula \citep{1986AIHS...44..263D,Wex14} rooted in the PN theoretical framework based on the DD parameterization \citep{1985AIHS...43..107D}, which is not used in Solar System studies.
In the case of the recent detection of the Schwarzschild precession (1PN) in the orbit of the S-star S2 \citep{2020A&A...636L...5G}, astrometric measurements of the stellar right ascension (RA) and declination (DEC), and spectroscopic observations of the star's radial velocity were used to extract it with a 14-parameter fit including, among other things, the six parameters of the orbit and a dimensionless scaling parameter $f_\mathrm{SP}$ of \rfr{konzo} which is $0$ for Newton and $1$ for GR.
Rather, \rfr{totKoPo} and \rfr{finis} should be interpreted, for a given astronomical system, as useful tools to quickly perform sensitivity analyses and error budgets starting from information usually available about it in order to preliminarily check if the 2PN effects are still far from the current or future experimental sensitivity, and if other competing dynamical features of motion are larger than them. The way the actual measure would ultimately be accomplished is entirely another matter which depends on the specific system at hand.
}
\section{Looking at some of the short-period S-stars in Sgr A$^\ast$}\lb{sec6}
Recently, \citet{2020ApJ...899...50P} discovered some short-period S-stars orbiting the candidate supermassive black hole (SMBH) in the Galactic Center (GC) at Sgr A$^\ast$ \citep{2008ApJ...689.1044G,2010RvMP...82.3121G}; their relevant orbital parameters are reported in Table\,\ref{tavola0}.
\begin{table}[!htb]
\caption{Key orbital parameters of some fast-revolving S-stars as per Table 2 of \citet{2020ApJ...899...50P}. Here, the semimajor axis $a$ is in milliparsec (mpc),  the orbital period $\Pb$ is in yr, while the eccentricity $e$ is dimensionless.
}\lb{tavola0}
\begin{center}
\begin{tabular}{|c|c|c|c|}
\hline
& & &     \\[\dimexpr-\normalbaselineskip+2pt]
star & $a\,\ton{\mathrm{mpc}}$ & $\Pb\,\ton{\mathrm{yr}}$ & $e$ \\
\hline
& & &   \\[\dimexpr-\normalbaselineskip+2pt]
S4714 & $4.079\pm 0.012$ & $12.0\pm 0.3$ & $0.985\pm 0.011$ \\
S62 & $3.588 \pm 0.02$ & $9.9\pm 0.3$ & $0.976\pm 0.01$ \\
S4711 & $3.002 \pm 0.06$ & $7.6\pm 0.3$ & $0.768\pm 0.030$  \\
\hline
\end{tabular}
\end{center}
\end{table}
Since some of them can move as fast as $\simeq 10$ per cent of the speed of light $c$ at perinigricon, it seems appropriate to look at their 2PN orbital effects. Lately, \citet{2020ApJ...901L..32F,2020ApJ...904..186I,2020ApJ...899...50P} investigated the 1PN gravitomagnetic Lense-Thirring orbital precessions, induced by the SMBH's angular momentum $\bds J_\bullet$, in view of a possible detection in future.

\begin{figure}[!htb]
\begin{center}
\centerline{
\vbox{
\begin{tabular}{c}
\epsfysize= 6.0 cm\epsfbox{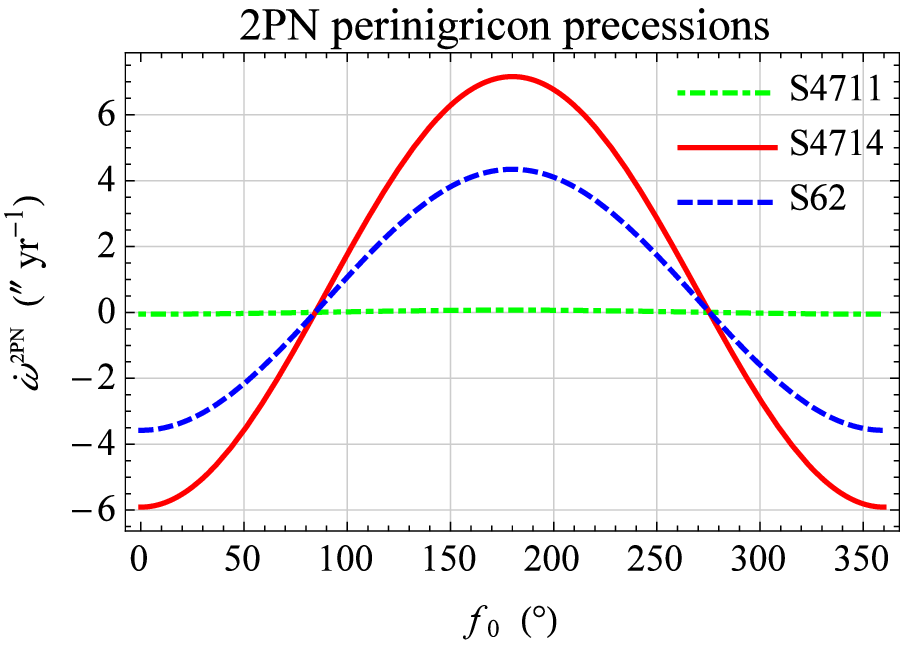}\\
\epsfysize= 6.0 cm\epsfbox{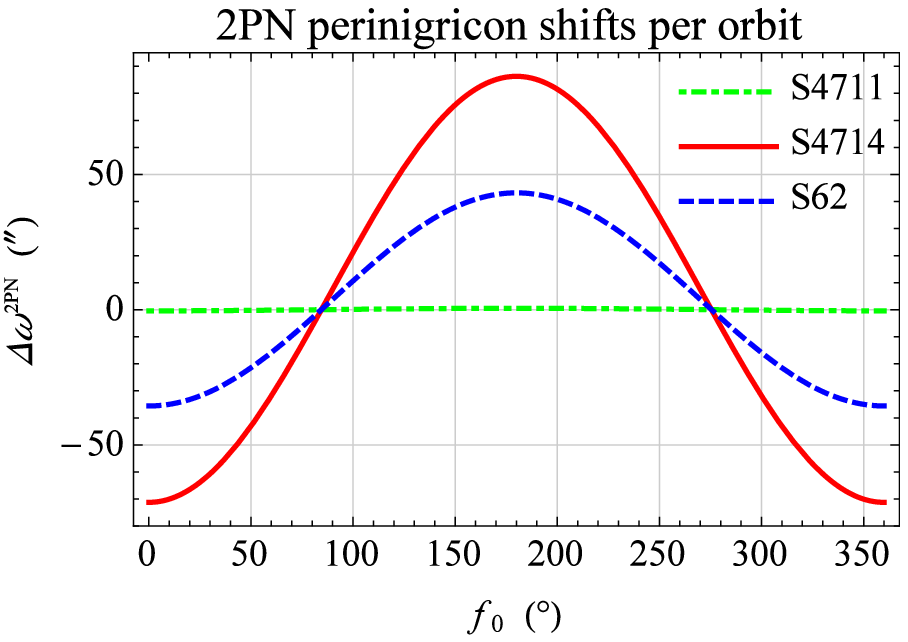}\\
\end{tabular}
}
}
\caption{Upper panel: plot of the 2PN perinigricon precessions $\dot\omega^\mathrm{2PN}$ of the S-stars S4711, S4714, S62 \citep{2020ApJ...899...50P}, in arcseconds per year $\ton{^{\prime\prime}\,\mathrm{yr}^{-1}}$,  as a function of the true anomaly $f_0$ at epoch according to \rfr{totKoPo}. Lower panel: plot of the 2PN perinigricon shifts per orbit $\Delta\omega^\mathrm{2PN}=2\uppi/\nk\,\dot\omega^\mathrm{2PN}$ of the S-stars S4711, S4714, S62 \citep{2020ApJ...899...50P}, in arcseconds $\ton{^{\prime\prime}}$,  as a function of the true anomaly $f_0$ at epoch according to \rfr{totKoPo}.}\label{figura1}
\end{center}
\end{figure}

Figure\,\ref{figura1} displays the 2PN perinigricon precessions $\dot\omega^\mathrm{2PN}$ and shifts per orbit $\Delta\omega^\mathrm{2PN}$ of the S-stars S4711, S4714, S62 as functions of $f_0$ according to \rfr{totKoPo}.
It turns out that the largest precessions, of the order of $\left|\dot\omega^\mathrm{2PN}\right|\lesssim 4-6\,\mathrm{arcseconds\,per\,year}\,\ton{^{\prime\prime}\,\mathrm{yr}^{-1}}$, occur for S62 and S4714, whose shifts per orbit can be as large as $\left|\Delta\omega^\mathrm{2PN}\right|\lesssim 40-90\,^{\prime\prime}$. Such figures are about of the same order of magnitude of the largest possible values of the 1PN Lense-Thirring precessions for $\omega$, as shown by Table\,4 of \citet{2020ApJ...904..186I}. Although future improvements in the observational techniques may possibly bring such post-Newtonian effects within the range of measurability \citep{2020ApJ...899...50P}, it should be remarked that they may be biased by the currently much larger systematic bias due to the lingering uncertainties in the stellar orbital parameters and the SMBH's mass entering the dominant 1PN component of the perinigricon precession \textcolor{black}{of \rfr{konzo}};
see Table\,5 of \citet{2020ApJ...904..186I}.
Moreover, as a further source of systematic bias, it is known that also an extended mass distribution around the SMBH could concurrently affect the perinigricon precession \citep{2001A&A...374...95R,2007PASP..119..349N,2007PhRvD..76f2001Z}.
\section{Summary and conclusions}\lb{sec7}
After having disclosed and corrected the calculational errors affecting \citet{2020Univ....6...53I}, it was demonstrated that the approaches by \citet{1988NCimB.101..127D,1994ARep...38..104K,2020Univ....6...53I} are, in fact,  equivalent in analytically calculating the 2PN pericentre precession. Indeed, they yield the same result once the appropriate conversions from the adopted parameterizations are made. This demonstrates that
the approach by \citet{2020Univ....6...53I} is correct.
%
%
%

The fact that the total 2PN pericentre rate of \rfr{totKoPo} contains also $f_0$ due to the indirect contributions to it arising from the cross-coupling of the 1PN acceleration with itself in the calculational procedure is an example of a general feature which seems to characterize also other mixed effects due to the interplay of different accelerations (Newtonian and PN) in the orbital averaging of the Gauss equations when the instantaneous shifts of all the Keplerian orbital elements are taken into account as well. The same feature occurs, e.g., also for the mixed precessions of the order of $J_2/c^2$ due to the Newtonian quadrupole acceleration and the 1PN one, as explicitly calculated also by other authors in the literature. Nonetheless, this fact should not induce one to think about it as a sign of some new physics, being simply a matter of choice of a particular gauge. Choosing a given coordinate system and a gauge is a standard and unavoidable practice in actual data reductions and in designing possible future tests with astronomical and astrophysical systems of potential interest.

To this aim, \rfr{totKoPo} was used to calculate the magnitude of the 2PN perinigricon precessions for some of the recently discovered fast-orbiting S-stars in Sgr A$^\ast$ whose orbital configuration is given in the literature just in terms of the osculating Keplerian orbital elements adopted here. It turns out that the shifts per orbit of S62 and S4714 can be as large as $\simeq 40-90$ arcseconds, being, in principle, detectable in view of the expected progress in the observational techniques. Nonetheless, a major source of systematic bias lies in the lingering uncertainty in the much larger 1PN component of the stellar perinigricon rates.
\section*{Acknowledgements}
I am grateful to M. Efroimsky for useful advices.
\bibliography{2PN}{}

\end{document}